\documentclass{appolb}
\usepackage{custom}

\begin{document}
\title{NON-GAUSSIAN FLUCTUATION DYNAMICS%
\thanks{Presented at the XXIXth International Conference on Ultra-relativistic Nucleus-Nucleus Collisions (Quark Matter 2022), April 4-10, 2022, Krak\'{o}w, Poland.}%
}

\author{Xin An
\address{Department of Physics and Astronomy, University of North Carolina, Chapel Hill, NC 27599, USA}
}


\maketitle
\begin{abstract}
Recent progress of a general deterministic approach to the non-Gaussian fluctuation dynamics is reviewed, with an emphasis on the derivation of the fluctuation evolution equations and their phenomenological implication in heavy-ion collision experiments.
\end{abstract}
  
\section{Introduction}
Fluctuations are ubiquitous phenomena emerging on all length scales, from the cosmological length all the way to the size of the quark-gluon plasma created in heavy-ion collisions, where they could potentially play essential roles. In many situations, fluctuations are resulted from addition of independent and random sources on smaller scales and are approximately described by the Gaussian distribution, per the central limit theorem. In addition, the time scale of system evolution is usually much larger than the time scale of the fluctuation equilibration process, so that fluctuations can also be well approximated as in equilibrium. Nonetheless, there are certain situations where fluctuations can significantly deviate from their Gaussian and/or equilibrium distribution, and these deviations arouse more and more interests on various physical subjects in recent years. A typical example, which will be focused in this work, is the event-by-event fluctuations of particle multiplicities in heavy-ion collision experiments. The shape of the multiplicity distribution for the observed particles (e.g., net protons), characterized by its cumulants, carries important information of the collision process. For example, it was argued that in equilibrium the higher-order cumulants are more sensitive to the critical point \cite{Stephanov:2008qz,Stephanov:2011pb}, which may give rise to the non-monotonic energy dependence of the particle fluctuations in heavy-ion collisions -- an intriguing hint of the QCD critical point \cite{PhysRevLett.126.092301}. However, since the fireballs, created by the collisions of two ultrarelativistic nuclei, are short lived and rapidly expanding, it would be natural to expect that the fluctuations may fall out of equilibrium. Thus the argument in Refs.~\cite{Stephanov:2008qz,Stephanov:2011pb} based on equilibrium theory needs to be amended to incorporate the non-equilibrium effects of non-Gaussian fluctuations -- the major motivation of Refs.~\cite{An_2021,An_2022b}. In this proceeding, the progress achieved in Ref.~\cite{An_2021} will be reviewed in Sec.~\ref{sec:theory}. Additional numerical results will be presented and discussed in Sec.~\ref{sec:example}. These works altogether constitute an integral part of the BEST framework \cite{An_2022} for interpreting the forthcoming experimental results from the second phase of the RHIC Beam Energy Scan program.

\section{Fluctuation evolution equations}
\label{sec:theory}

The {\em Langevin equations} for a set of stochastic fields $\psi_i$'s, with the index $i=(i_1,i_2,\dots)$ labeling both the variables and their associated space coordinates (if applicable), read
\begin{equation}\label{eq:Langevin}
    \partial_t\psi_i = F_i + \xi_i
\end{equation}
where $F_i$ and $\xi_i$ are the drift force and random force (noise) respectively. The noise is assumed to be Gaussian, i.e.,
\begin{equation}\label{eq:noise}
    \langle\xi_{i_1}(t_1)\xi_{i_2}(t_2)\rangle=2Q_{i_1i_2}\delta(t_1-t_2)\,,
\end{equation}
and its amplitude $Q_{ij}$, known as the Onsager matrix, is set by the fluctuation-dissipation relation. One can convert the Langevin equations \eqref{eq:Langevin} into the {\em Fokker-Planck equation} for the probability distribution function of the stochastic fields, $P(\psi_i;t)$ \cite{ZinnJustin:2002ru}:
\begin{equation}\label{eq:FP}
   \partial_t P=\left(-F_i P + \left(Q_{ij}P\right)_{,\,j}\right)_{,\,i}
\end{equation}
where $(\dots)_{,i}=\delta(\dots)/\delta\psi_i$.\footnote{This equation is invariant under $Q_{ij}\to Q_{ij}+\Omega_{ij}$ where $\Omega_{ij}=-\Omega_{ji}$ is the antisymmetric matrix obtained from the Poisson bracket of stochastic fields \cite{ZinnJustin:2002ru}.} The Fokker-Planck equation solves two problems the Langevin equation suffered from: the problem of {\em infinite noise} due to the divergence of $\delta$-function in Eq.~\eqref{eq:noise} when $t_1\to t_2$; and the problem of {\em multiplicative noise} due to the ambiguity of the noise amplitude $Q_{ij}$ in the continuous limit $\Delta t\to0$ where $\Delta t$ is the discretized time increment of the evolution. The divergence in the first problem can be regularized analytically using the renormalization techniques \cite{An:2019rhf,An:2019fdc,An:2020jjk}, while the ambiguity in the second problem can be eliminated by specifying the discretization prescription, e.g., Ito's prescription as in Eq.~\eqref{eq:FP}.

\begin{figure}[htb]
\centerline{%
\includegraphics[width=16.2cm]{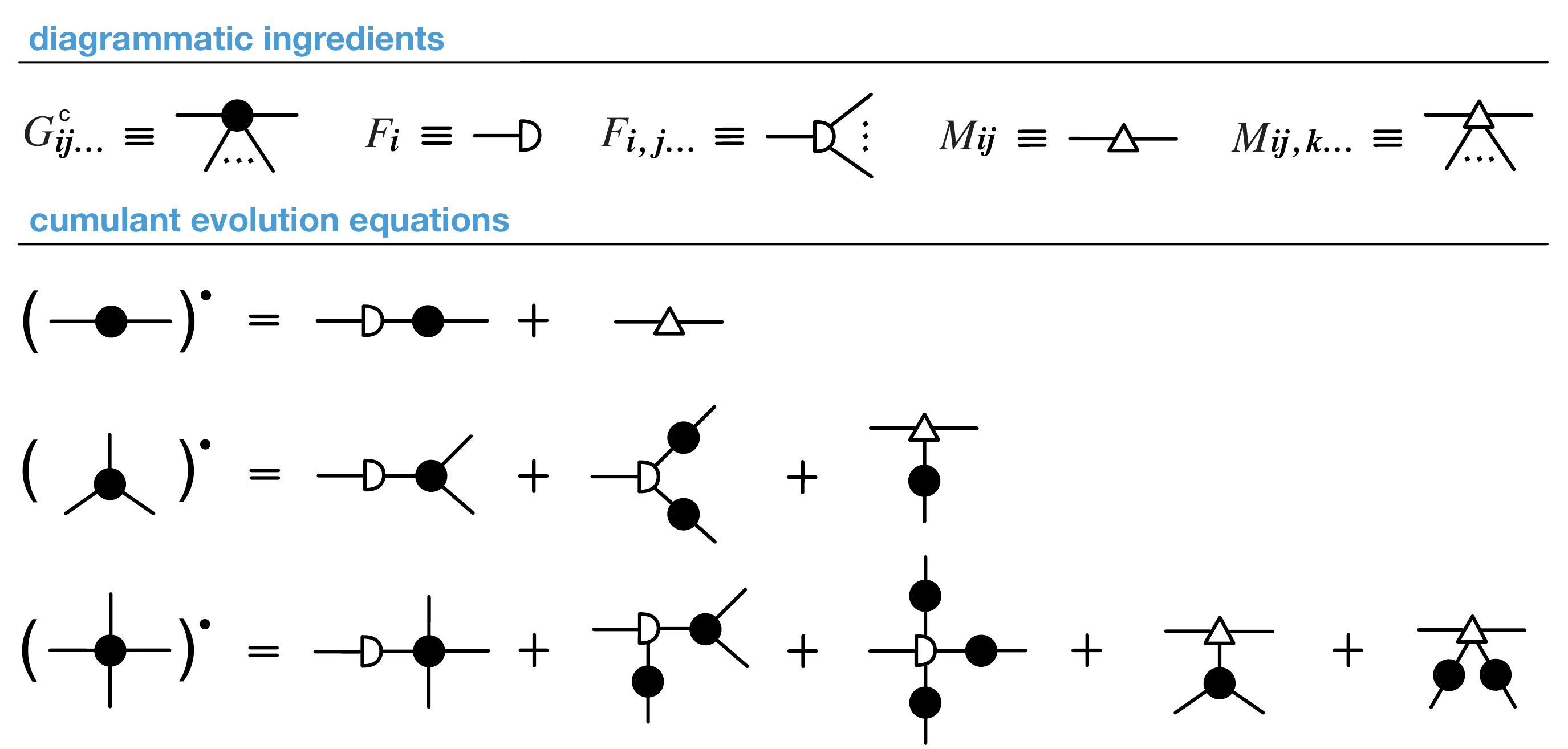}
}
\caption{\label{fig:diagrams}Diagrammatic representation of Eq.~\eqref{eq:dtGc} (or Eqs.~\eqref{eq:dtW_charge}) for $n=2,3,4$ in terms of the diagrammatic ingredients introduced in this figure. The dot on the left hand side denotes the time derivative $\partial_t$.}
\end{figure}

The cumulant generating functional, $\mathcal W[\mu_i]$, is defined through
\begin{equation}\label{eq:Z}
   e^{\mathcal W[\mu_i]}=\langle e^{\mu_i\psi_i}\rangle\equiv\int\mathcal D\psi Pe^{\mu_i\psi_i}\,,
\end{equation}
which can be expanded in a series of the external sources $\mu_i$'s: 
\begin{equation}\label{eq:W}
   \mathcal W=\sum_{n=1}^\infty\frac{1}{n!}G^{\rm c}_{i_1\dots i_n}\mu_{i_1}\dots\mu_{i_n}
\end{equation}
where the expansion coefficient $G^{\rm c}_{i_1\dots i_n}\equiv G^{\rm c}_n\equiv\delta\mathcal W/\delta\mu_{i_1}\dots\delta\mu_{i_n}|_{\mu=0}$ is the $n$-th cumulant. Using Eqs.~\eqref{eq:FP} and \eqref{eq:Z} one arrives at the evolution equation for $\mathcal W[\mu_i]$: 
\begin{equation}\label{eq:dtW}
   \partial_t \mathcal W=e^{-\mathcal W}\left(\mu_iF_i+\mu_i\mu_jQ_{ij}\right)e^{\mathcal W}
\end{equation}
where $F_i=F_i(\delta/\delta\mu_i)$, $Q_{ij}=Q_{ij}(\delta/\delta\mu_i)$. Substituting Eq.~\eqref{eq:W} into \eqref{eq:dtW} one can readily obtain the evolution equation for cumulant $G^{\rm c}_n$. However, these equations are not in closed form, i.e., the equations for lower-order cumulants also depend on higher-order cumulants. Fortunately, in certain regimes where some parameters are legitimately small, these equations can be systematically truncated. We introduced two such parameters. The parameter $\varepsilon$, which is inversely proportional to the uncorrelated degrees of freedom, controls the loop expansion. Another parameter $\delta$, identified as the Knudsen number in hydrodynamics, controls the gradient expansion. Although each of them plays its own role, they are not independent and are indeed related by $\varepsilon\sim\delta^3$ in hydrodynamics. Assigning the following power counting:
\begin{equation}
   F_i\sim\delta^2\,, \qquad Q_{ij}\sim\delta^2\varepsilon\,, \qquad G^{\rm c}_n\sim\varepsilon^{n-1}\,,
\end{equation}
and keeping the leading terms ($\sim\delta^2\varepsilon^{n-1}$) that turn out to be connected tree-level diagrams, we arrive at the truncated and iteratively solvable equations, schematically formulated in the form
\begin{equation}\label{eq:dtGc}
   \partial_t G^{\rm c}_n=\mathscr F_{\rm tree}[\{G^{\rm c}_n, G^{\rm c}_{n-1},\dots,G^{\rm c}_2,\langle\psi\rangle\}]
\end{equation}
and diagrammatically represented in Fig.~\ref{fig:diagrams}. Their explicit forms can be found in Refs.~\cite{An_2021,An_2022b}.

If the variables $\psi_{i_1},\dots,\psi_{i_n}$ are fields defined locally in continuous real space at $\bm x_1,\dots,\bm x_n$, it may be more convenient to analyze Eq.~\eqref{eq:dtGc} in the wave-vector space. To this end we introduced a novel {\em multi-point} Wigner function \cite{An_2021}:
\begin{equation}\label{eq:Wc}
    W^{\rm c}_{n}(\bm x,\bm q_1,\dots,\bm q_n)=\bigintsss\left[\prod_{i=1}^n d^3\bm y_ie^{-i\bm q_i\cdot\bm y_i}\right]\delta^{(3)}\left(\frac{1}{n}\sum_{i=1}^n\bm y_i\right)G^{\rm c}_{n}(\bm x_1,\dots,\bm x_n)\,,
\end{equation}
where $\bm q_i$ is the wave-vector conjugate to $\bm y_i\equiv\bm x_i-\bm x$ and $\bm x=\sum_{i=1}^n\bm x_i/n$. Under this Wigner transform we imposed the constraint $\sum_{i=1}^n\bm y_i=0$, as a consequence $\bm q_i$'s are not independent, and must also sum to zero after an appropriate shift. Obviously, Eq.~\eqref{eq:Wc} reproduces the traditional definition of Winger function when $n=2$. Using Eq.~\eqref{eq:Wc} one can represent Eq.~\eqref{eq:dtGc} in the wave-vector space. We defer its presentation in a specific problem discussed in Sec.~\ref{sec:example}. 

For {\em confluent} formulation of non-Gaussian fluctuations in relativistic fluid, see Ref.~\cite{An_2022b}.

\section{Non-Gaussian fluctuation dynamics of diffusive charge}
\label{sec:example}

In this section we apply our general formalism to a specific problem -- the evolution of diffusive charge, using the translation given by Table.~\ref{tab:dictionary}. In this problem the stochastic variable, charge density $n(\bm x)$, is defined in continuous space, the Onsager matrix $Q_{ij}$ is related to conductivity $\lambda$ through the fluctuation-dissipation relation, and the drift force $F_i$ is given by the divergence of a diffusion current whose constitutive relation, $\lambda\bm\nabla\alpha$ where $\alpha$ is the chemical potential per temperature, can be determined by the second law of thermodynamics.

  \newcommand\Tstrut{\rule{0pt}{2.9ex}}         
     \newcommand\Bstrut{\rule[-1.4ex]{0pt}{0pt}}   
     \setlength{\arrayrulewidth}{0.3mm}
     \setlength{\tabcolsep}{42pt}
    \begin{table}[htb]
    \begin{tabular}{ c |c | c }
    \hline \hline
            quantities & general & diffusive charge  \Tstrut\Bstrut\\ 
            \hline
            variable & $\psi_i$ & $n(\bm x)$ \Tstrut\Bstrut\\
            \hline
            variable index & $i,j,k,$ etc. & $\bm x,\bm y, \bm z,$ etc. \Tstrut\Bstrut\\ 
            \hline
            Onsager matrix & $Q_{ij}$ & $\bm\nabla_{\bm x} \lambda \bm\nabla_{\bm y}\,\delta^{(3)}_{\bm{xy}}$ \Tstrut\Bstrut\\
            \hline
            drift force & $F_{i}$ & $\bm\nabla_{\bm x}\lambda\bm\nabla_{\bm x}\alpha$ \Tstrut\Bstrut\\
        \hline \hline
    \end{tabular}
    \caption{\label{tab:dictionary}Translation of general formalism to the problem of diffusive charge.}
    \end{table}

Applying the translation in Table.~\ref{tab:dictionary} to Eq.~\eqref{eq:dtGc} and using Eq.~\eqref{eq:Wc}, one immediately obtains the evolution equations for the diffusive charge cumulants in the wave-vector space. The first few equations for $n=2,3,4$ read (cf. Fig.~\ref{fig:diagrams}):
\begin{align}\label{eq:dtW_charge}
   \partial_tW_2(\bm q_1)=&-2\left[\gamma\bm q_1^2W_2(\bm q_1)+\lambda\bm q_1\cdot\bm q_2\right]_{\overline{12}}\,,\nn
   \partial_t W_3(\bm q_1,\bm q_2)=&-3\left[\gamma\bm q_1^2W_3(\bm q_1,\bm q_2) + \gamma'\bm q_1^2W_2(\bm q_2)W_2(\bm q_3)+2\lambda'\bm q_1\cdot\bm q_2W_2(\bm q_3)\right]_{\overline{123}}\,,\nn
   \partial_t W^{\rm c}_4(\bm q_1,\bm q_2,\bm q_3)=&-4\left[\gamma\bm q_1^2W^{\rm c}_4(\bm q_1,\bm q_2,\bm q_3) + 3\gamma'\bm q_1^2W_2(\bm q_2)W_3(q_3,\bm q_4)+\gamma''\bm q_1^2W_2(\bm q_2)W_2(\bm q_3)W_2(\bm q_4)\right.\nn
   &\left.+3\lambda'\bm q_1\cdot\bm q_2W_3(\bm q_3,\bm q_4)+3\lambda''\bm q_1\cdot\bm q_2W_2(\bm q_3)W_2(\bm q_4)\right]_{\overline{1234}}\,,
\end{align}
where $\gamma=\lambda\alpha'$, $\alpha'\equiv\partial\alpha/\partial n$ and $\overline{1\dots n}$ denotes the sum over all $n!$ permutations of $\bm q_1,\dots,\bm q_n$ divided by $n!$. Eqs.~\eqref{eq:dtW_charge} are solved by $W_2^{\rm eq}=1/\alpha'$, $ W_3^{\rm eq}=-\alpha''/\alpha'^3$, $W_4^{\rm c,eq}=(3\alpha''^2-\alpha'\alpha''')/\alpha'^5$ in equilibrium, as expected from thermodynamic calculations.

In the critical regime where the correlation length $\xi$ is still much less than the fluctuation scale $q^{-1}\equiv|\bm q|^{-1}$ but becomes much larger than all other microscopical lengths such as the inverse of temperature, i.e., $T^{-1}\ll\xi\ll q^{-1}$, all thermodynamic and transport quantities will acquire their dependence on the correlation length (which serves as a UV cutoff of fluctuations). In the dynamical universal class of Model H, we have approximately \cite{An_2022b}
\begin{equation}\label{eq:xi}
   \alpha^{(n)}\sim\xi^{\frac{n-5}{2}}\,, \qquad \lambda^{(n)}\sim\xi^{\frac{n+2}{2}}\,, \qquad \gamma^{(n)}\sim\xi^{\frac{n-2}{2}}\,, \qquad W^{\rm c}_n\sim\xi^{\frac{5n-6}{2}}\,, \qquad \partial_tW^{\rm c}_n\sim q^2\xi^{\frac{5n-8}{2}}\,.
\end{equation}
The last relation in Eq.~\eqref{eq:xi} says that, all terms presented in each equation of \eqref{eq:dtW_charge} are in the same power of correlation length, demonstrating their equal importance near the critical point. 

\begin{figure}[htb]
\centerline{%
\includegraphics[width=7.8cm]{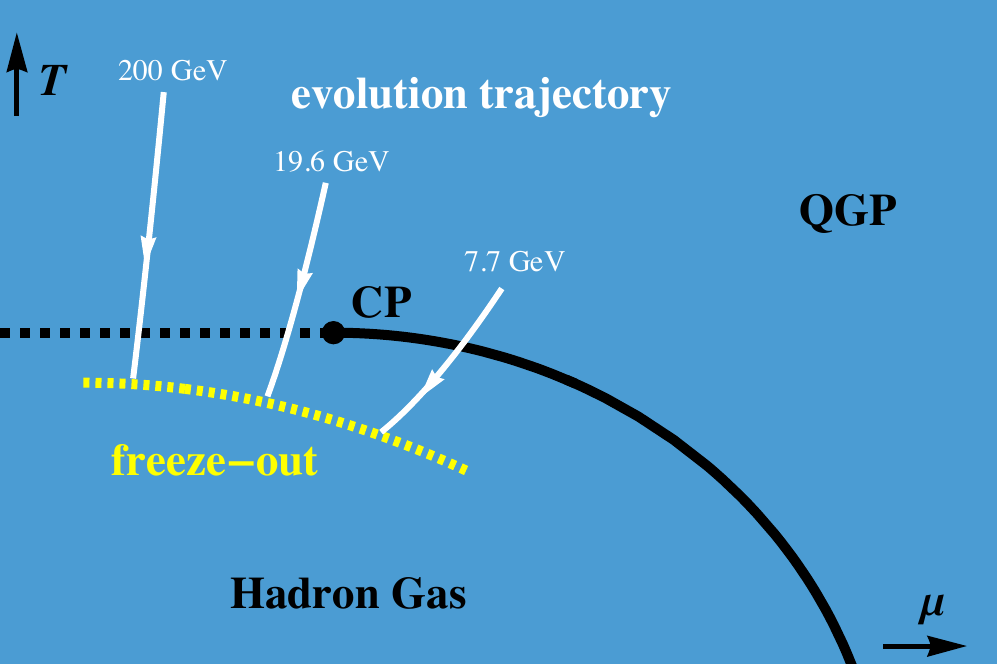}\hspace{0.1cm}\includegraphics[width=7.8cm]{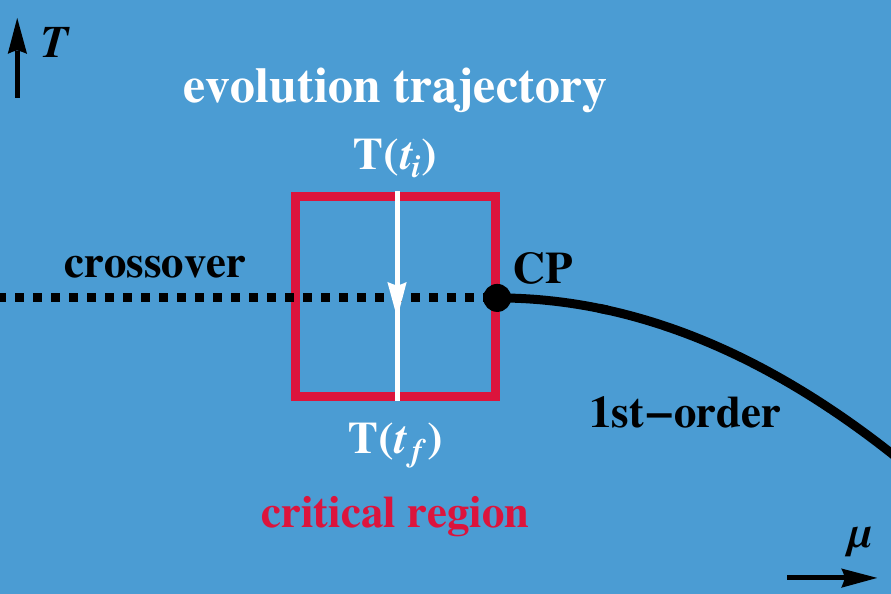}
}
\caption{\label{fig:qcd-cp}QCD phase diagram (left) and the zoomed-in critical region near the QCD critical point (right). The black dashed and solid curve is the crossover and first-order phase transition line. The white lines represent the evolution trajectories of each collision event mapped to the QCD phase diagram, with arrows indicating the time evolution direction. The yellow dashed curve is the freeze-out line where observables are measured. The critical region studied in Figs.~\ref{fig:ContourW} and \ref{fig:ContourW4} is bounded by the red square.}
\label{Fig:QCDphase}
\end{figure}

To see how Eqs.~\eqref{eq:dtW_charge} work in practice, let's consider the problem of baryon charge cumulant evolution in the crossover region near the QCD critical point (see Fig.~\ref{fig:qcd-cp}). Our purpose would be illustrating the robust phenomenological consequence of Eqs.~\eqref{eq:dtW_charge}. To this end, we postulate the following assumptions which simplify our simulations without losing the main features of the results. First, when establishing the dependence of the correlation length $\xi$ on baryon chemical potential $\mu$ and temperature $T$, we use a mapping $\xi(\mu,T)=\xi_{\rm Ising}(r(\mu,T),h(\mu,T))$ from Ising model where $r$ is the Ising temperature and $h$ is the Ising magnetic field, and assume the mapping is linear and orthogonal around the critical point $(\mu_{\rm c}, T_{\rm c})$, i.e., $\mu-\mu_{\rm c}\sim r$ and $T-T_{\rm c}\sim h$. Second, the evolution trajectories, rather complicated in reality, are assumed to be perpendicular to the crossover line (see the right panel of Fig.~\ref{fig:qcd-cp}). Third, the system is assumed to be in equilibrium at the initial time $t_i$, i.e., the moment when the system just enters the critical region, and assumed to freeze out before the final time $t_f$, i.e., the moment when the system leaves the critical region. Fourth, the system is assumed to be isotropic such that $W^{\rm c}_n$'s depend only on $n(n-1)/2$ independent scalar invariants, $\bm q_i\cdot\bm q_j$, $i<j$, which are all set to $q^2$ for simplicity \cite{An_2021}.

\begin{figure}[htb]
\centerline{%
\includegraphics[width=14.2cm]{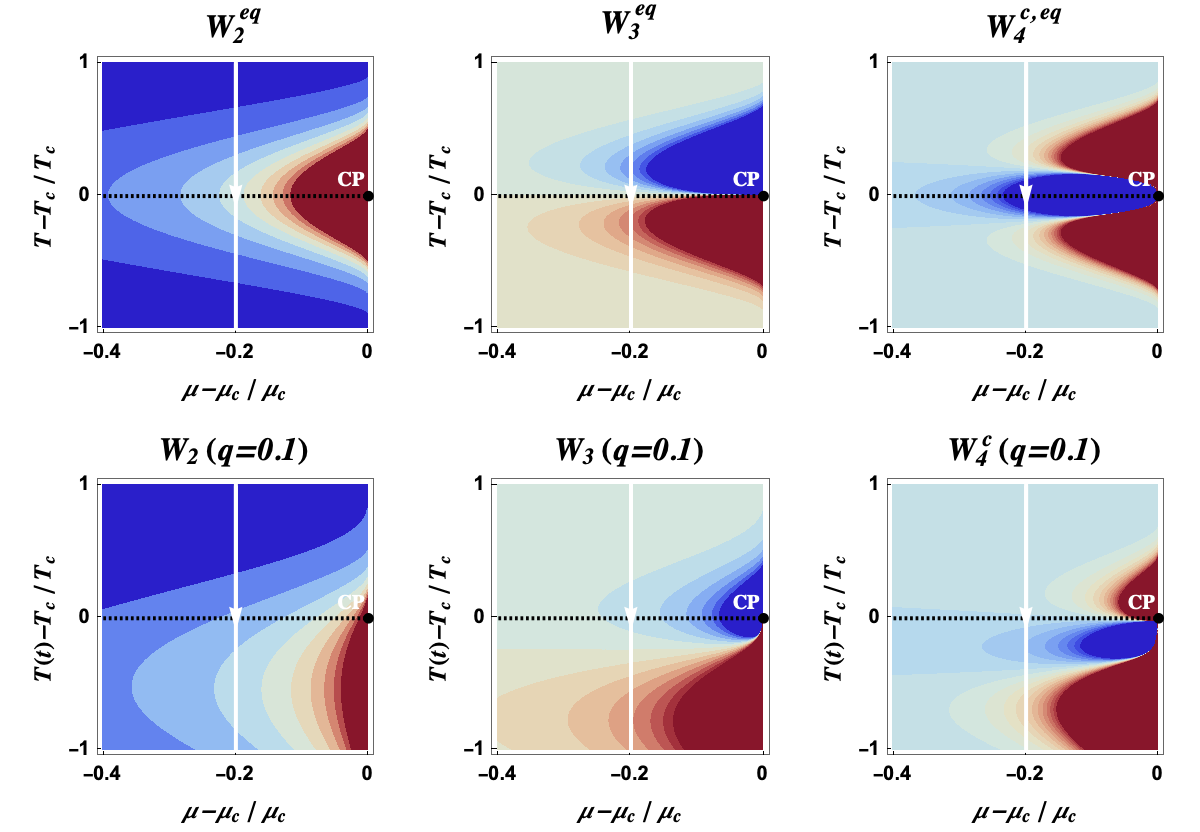}
}
\caption{\label{fig:ContourW}Contour maps of cumulants $W^{\rm c}_{n=2,3,4}$ in equilibrium (top panel, $q\to\infty$) and out of equilibrium (bottom panel, $q=0.1$).}
\end{figure}

Simulating Eqs.~\eqref{eq:dtW_charge} using the above assumptions, we find the contour maps of cumulants $W^{\rm c}_{n=2,3,4}$ in the critical region, shown in Fig.~\ref{fig:ContourW}. The top panel shows the contours of cumulants in equilibrium, which are symmetric to the crossover line due to aforementioned simplifications. The bottom panel shows a particular example (i.e., $q=0.1$)\footnote{Here we used arbitrary units, which are not specified thereby.} where the cumulants are out of equilibrium. In this case the contours are significantly distorted as if they are dragged toward the time evolution direction. In other words, although the dynamic cumulants keep approaching their instantaneous equilibrium values, they still retain memories of their past evolution history.\footnote{Similar study can be found in Ref.~\cite{Mukherjee:2015swa} for the homogeneous mode that is independent of the scale $q$.}

\begin{figure}[htb]
\centerline{%
\includegraphics[width=14cm]{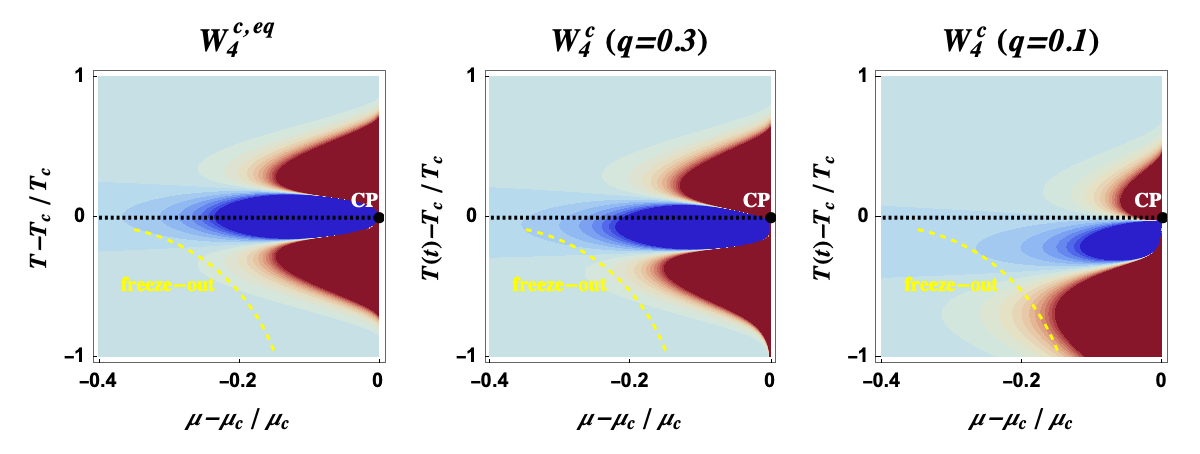}
}
\caption{\label{fig:ContourW4}Contour maps of fourth cumulant $W^{\rm c}_4$ at $q\to\infty$ (left), $q=0.3$ (middle) and $q=0.1$ (right) in the critical region. The freeze-out line (yellow dashed curve) is arbitrarily chosen for illustration purpose.}
\end{figure}

Focusing on the fourth cumulant $W^{\rm c}_4$, we plot the contours for different $q$'s in Fig.~\ref{fig:ContourW4}. We find that cumulants are closer to equilibrium on smaller scale (e.g., $q=0.3$), while retain a longer memory of past on larger scale (e.g., $q=0.1$). This is a consequence of causality: on larger spatial scale, fluctuations need longer time to equilibrate. When $q\to0$ (infinitely large scale), fluctuations will be completely suppressed due to the conservation of charges, and remains as if they were at initial time (which, of course, are also suppressed).

\begin{figure}[htb]
\centerline{%
\includegraphics[width=9cm]{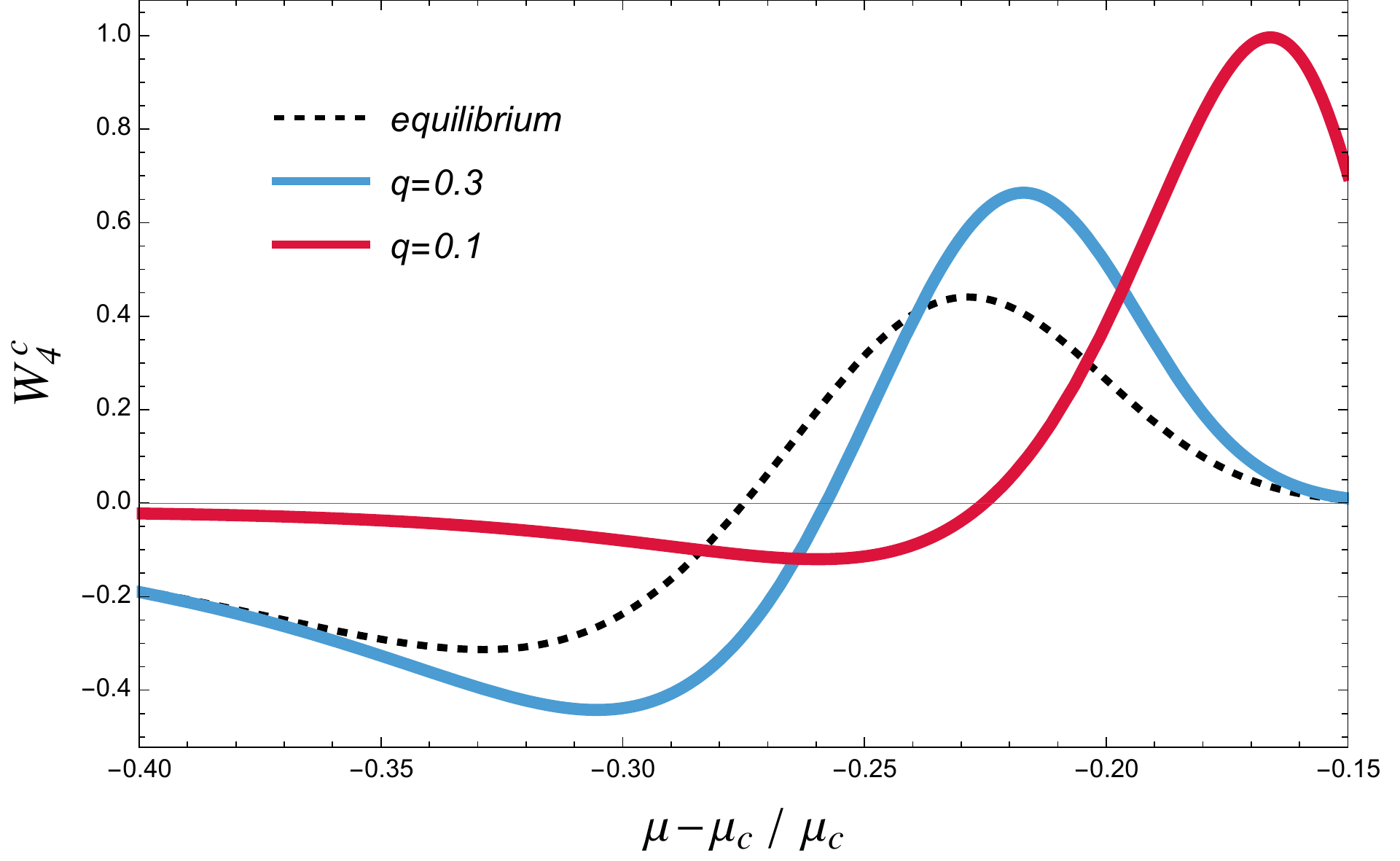}
}
\caption{\label{fig:FreezeoutW4} The dependence of $W^{\rm c}_4$ on baryon chemical potential $\mu$ along the freeze-out line in Fig.~\ref{fig:ContourW4}, at $q\to\infty$ (equilibrium), $q=0.3$ and $q=0.1$ respectively.}
\end{figure}

The fourth cumulant $W^{\rm c}_4$ is of particular interests to us, since its equilibrium value possesses more sensitivity to the critical point compare to the lower-order cumulants. If one reads off its value along the freeze-out line (yellow dashed curve) in Fig.~\ref{fig:ContourW4}, one finds its expected non-monotonicity as function of baryon chemical potential in Fig.~\ref{fig:FreezeoutW4}. Noting that the baryon chemical potential is inversely related to the collision energy, Fig.~\ref{fig:FreezeoutW4} is qualitatively consistent with the experiment data reported in Ref.~\cite{PhysRevLett.126.092301}. More interestingly, one finds that the curve for smaller $q$ (red colored) might be shifted toward large baryon chemical potential more significantly compared to the one for larger $q$ (blue colored). That is to say, the non-equilibrium effect, largely coming from the fluctuations with wave-vectors around $q\sim0.1-0.3$, might shift the curve predicted by equilibrium theory toward the lower collision energy. As a reminder, one should not compare the values here quantitatively, as our results are only for illustration purpose.

\section{Summary}

In this proceeding, we reviewed a general deterministic approach to non-Gaussian fluctuation dynamics. This novel approach is established with controllable perturbative loop/gradient expansion and multi-point Wigner function. In this approach we derived the general evolution equations for cumulants, and demonstrated their robust phenomenological implication via the problem of diffusive charge near the QCD critical point. Our numerical results suggest that the non-equilibrium effect to cumulants needs to be taken into account for a rigorous quantitative comparison with the experimental results. As for the future, the cumulant evolution equations need to be implemented in a more realistic setup, and our formalism is yet to incorporate all non-Gaussian hydrodynamic fluctuations with their freeze-out prescriptions \cite{An_2022b,An:2019rhf,An:2019fdc,Pradeep:2022mkf}.

\bibliographystyle{utphys}
\bibliography{references}

\providecommand{\href}[2]{#2}\begingroup\raggedright\begin{thebibliography}{10}

\bibitem{Stephanov:2008qz}
M.~Stephanov, ``{Non-Gaussian fluctuations near the QCD critical point},''
  \href{http://dx.doi.org/10.1103/PhysRevLett.102.032301}{{\em Phys. Rev.
  Lett.} {\bfseries 102} (2009) 032301},
\href{http://arxiv.org/abs/0809.3450}{{\ttfamily arXiv:0809.3450 [hep-ph]}}.

\bibitem{Stephanov:2011pb}
M.~Stephanov, ``{On the sign of kurtosis near the QCD critical point},''
  \href{http://dx.doi.org/10.1103/PhysRevLett.107.052301}{{\em Phys. Rev.
  Lett.} {\bfseries 107} (2011) 052301},
\href{http://arxiv.org/abs/1104.1627}{{\ttfamily arXiv:1104.1627 [hep-ph]}}.

\bibitem{PhysRevLett.126.092301}
{\bfseries STAR} Collaboration, J.~Adam {\em et~al.}, ``Nonmonotonic energy
  dependence of net-proton number fluctuations,''
  \href{http://dx.doi.org/10.1103/PhysRevLett.126.092301}{{\em Phys. Rev.
  Lett.} {\bfseries 126} (2021) 092301},
  \href{http://arxiv.org/abs/2001.02852}{{\ttfamily arXiv:2001.02852
  [nucl-ex]}}.

\bibitem{An_2021}
X.~An, G.~Ba{\c{s}}ar, M.~Stephanov, and H.-U. Yee, ``Evolution of non-gaussian
  hydrodynamic fluctuations,''
  \href{http://dx.doi.org/10.1103/PhysRevLett.127.072301}{{\em Phys. Rev.
  Lett.} {\bfseries 127} (2021) 072301},
  \href{http://arxiv.org/abs/2009.10742}{{\ttfamily arXiv:2009.10742
  [hep-th]}}.

\bibitem{An_2022b}
{X. An, G. Ba{\c{s}}ar, M. Stephanov, and H.-U. Yee}, ``{Non-Gaussian
  fluctuation dynamics in relativistic fluid}.'' to appear.

\bibitem{An_2022}
X.~An {\em et~al.}, ``The {BEST} framework for the search for the {QCD}
  critical point and the chiral magnetic effect,''
  \href{http://dx.doi.org/10.1016/j.nuclphysa.2021.122343}{{\em Nucl. Phys. A}
  {\bfseries 1017} (2022) 122343},
  \href{http://arxiv.org/abs/2108.13867}{{\ttfamily arXiv:2108.13867
  [nucl-th]}}.

\bibitem{ZinnJustin:2002ru}
J.~Zinn-Justin, ``{Quantum field theory and critical phenomena},'' {\em Int.
  Ser. Monogr. Phys.} {\bfseries 113} (2002) 1--1054.

\bibitem{An:2019rhf}
X.~An, G.~Ba\c{s}ar, M.~Stephanov, and H.-U. Yee, ``{Relativistic Hydrodynamic
  Fluctuations},'' \href{http://dx.doi.org/10.1103/PhysRevC.100.024910}{{\em
  Phys. Rev. C} {\bfseries 100} (2019) 024910},
\href{http://arxiv.org/abs/1902.09517}{{\ttfamily arXiv:1902.09517 [hep-th]}}.

\bibitem{An:2019fdc}
X.~An, G.~Ba\c{s}ar, M.~Stephanov, and H.-U. Yee, ``{Fluctuation dynamics in a
  relativistic fluid with a critical point},''
  \href{http://dx.doi.org/10.1103/PhysRevC.102.034901}{{\em Phys. Rev. C}
  {\bfseries 102} (2020) 034901},
  \href{http://arxiv.org/abs/1912.13456}{{\ttfamily arXiv:1912.13456
  [hep-th]}}.

\bibitem{An:2020jjk}
X.~An, ``{Relativistic dynamics of fluctuations and QCD critical point},''
  \href{http://dx.doi.org/10.1016/j.nuclphysa.2020.121957}{{\em Nucl. Phys. A}
  {\bfseries 1005} (2021) 121957},
  \href{http://arxiv.org/abs/2003.02828}{{\ttfamily arXiv:2003.02828
  [hep-th]}}.

\bibitem{Mukherjee:2015swa}
S.~Mukherjee, R.~Venugopalan, and Y.~Yin, ``{Real time evolution of
  non-Gaussian cumulants in the QCD critical regime},''
  \href{http://dx.doi.org/10.1103/PhysRevC.92.034912}{{\em Phys. Rev. C}
  {\bfseries 92} (2015) 034912},
  \href{http://arxiv.org/abs/1506.00645}{{\ttfamily arXiv:1506.00645
  [hep-ph]}}.

\bibitem{Pradeep:2022mkf}
M.~Pradeep, K.~Rajagopal, M.~Stephanov, and Y.~Yin, ``{Freezing out
  fluctuations in Hydro+ near the QCD critical point},''
  \href{http://arxiv.org/abs/2204.00639}{{\ttfamily arXiv:2204.00639
  [hep-ph]}}.

\end{thebibliography}\endgroup

\end{document}